\documentclass[10pt,aps,prl,twocolumn,groupedaddress,showkeys]{revtex4-1}
\usepackage{graphicx}
\usepackage{amsmath}
\usepackage{xcolor}
\usepackage{epstopdf}

\usepackage{hyperref}

\begin{document}

\title{Optoelectronic mixing with high frequency graphene transistors}
\author{A, Montanaro$^{1}$}\thanks{These authors contributed equally}
\author{W. Wei$^{2}$}\thanks{These authors contributed equally}
\author{D. De Fazio$^3$}
\author{U. Sassi$^3$}
\author{G. Soavi$^3$}
\author{A. C. Ferrari$^3$}
\author{H. Happy$^2$}
\author{P. Legagneux$^1$}
\author{E. Pallecchi$^2$}
\affiliation{$^1$Thales Research and Technology, 1, Avenue Augustin Fresnel, 91767 Palaiseau, France}
\affiliation{$^2$IEMN, University of Lille, CNRS UMR8520, Avenue Poincare CS-60069, 59652 Villeneuve d'Ascq, France}
\affiliation{$^3$Cambridge Graphene Centre, University of Cambridge, Cambridge CB3 0FA, UK}

\begin{abstract}

Graphene is ideally suited for optoelectronic applications. It offers absorption at telecom wavelengths, high-frequency operation and CMOS-compatibility. We report optoelectronic mixing up to to 67GHz using a back-gated graphene field effect transistor (GFET). We also present a model to describe the resulting mixed current. These results pave the way for GETs optoelectronic mixers for mm-wave applications, such as telecommunications and RADAR/LIDAR systems.
\end{abstract}
\maketitle
\section{Introduction}
Mixers are a key component of modern communication modules\cite{Maas1986}. In telecommunications and in radio detection and ranging (RADAR) systems\cite{Gagl1976}, the receiver analyzes the modulation of a carrier wave (or waveform) with frequencies in the microwave (3-30GHz) or mm-wave (30-300GHz) range, to extract information\cite{Gu2006,Skol2001}. Since signal processing is performed at near-zero frequencies (baseband\cite{Gu2006,Skol2001}), frequency downconversion is required\cite{Gu2006,Skol2001}. Downconversion is performed by mixing the modulated high frequency signal centered around the RF carrier frequency $f_{RF}$ with a local oscillator signal at frequency $f_{LO}$. This translates the modulation centered around $f_{RF}$ to $f_{IF}=f_{LO}-f_{RF}$. The local oscillator frequency is typically set near $f_{RF}$ so that $f_{IF}$ is close to zero\cite{Gu2006,Skol2001}. Superheterodyne receivers are a common type of radio receivers using frequency downconversion to process the original signal\cite{Poza2009}. For multi-antenna systems, it is preferable to use a single optical signal as a local oscillator and distribute it to each antenna\cite{ChizERC2014}, decreasing the receiver complexity and noise. For this purpose, one option is to use photodetectors (PDs) to transfer the local oscillator signal from the optical to the electrical domain\cite{ChizERC2014}. For this, an electrical mixer is used\cite{Poza2009}. A second option is to employ optoelectronic mixers\cite{RuffAS2000}, $\textit{i.e.}$ PDs capable of directly mixing the optical local oscillator with an electrical signal\cite{RuffAS2000}. Optoelectronic mixers (OEMs) are particularly convenient in RADAR and light detection and ranging (LIDAR) applications\cite{PillEORS2008,RuffAS2000,GhelN2014,VercOL2015}. State of the art OEMs at 1.55$\mu$m are based on III-V semiconductors epitaxially grown on InP\cite{ChoiIEEE2015,RouvOE2011}. These are efficient, but expensive, and can only be heterogeneously integrated in a Si platform\cite{ChoiIEEE2015,RouvOE2011}. Low cost and complementary metal-oxide-semiconductor (CMOS) compatible OEMs require CMOS compatible materials absorbing light at 1.55$\mu$m\cite{Sze2006}.

Graphene is promising for optoelectronics\cite{BonaNP2010,KoppNN2014,RomaNRM2018,SunACS2010}, with reported mobilities up to$\sim150000\text{ cm}^2\text{V}^{-1}\text{s}^{-1}$ at room temperature\cite{PurdNC2018}, a short ($\sim$1ps) photocarrier lifetime\cite{BreuPRB2011,BridNC2013,TomaPRB2013,LazzPRB2008} and a 2.3\%  broadband light absorption (including telecom wavelength)\cite{NairS2008}. Graphene-based optoelectronic devices are compatible with Si technology platforms\cite{PospNP2013}. Therefore, graphene-based optoelectronics mixers could combine telecom operation and CMOS compatibility.

The first high frequency optoelectronic mixer based on graphene was reported in Ref.\citenum{MontNL2016}. It was based on a coplanar waveguide integrating a graphene channel (GCPW) grown by chemical vapor deposition (CVD). The RF signal was injected into the GCPW while a 1.55$\mu$m laser illuminated the channel. Optoelectronic mixing was based on the linear dependence of the photocurrent on both optical incident power (P$_{opt}$) and voltage drop (V$_{bias}$) along the channel. As the photocurrent is proportional to P$_{opt}$·V$_{bias}$, upconverted and downconverted signals were generated. This GCPW operated up to 30GHz, with a conversion efficiency (\emph{i.e.} ratio of output power at $f_{IF}$ and input power at $f_{RF}$) of -85dB for a 10GHz modulated signal. These results are far from state-of-the-art OEMs performances achieved with III-V semiconductor-based Uni-Traveling Carrier Photodiodes (UTCPDs): -22 dB conversion efficiency at 35 Hz\cite{MohaOE2018}, and -40dB at 100GHz\cite{RouvOE2011}. However, the CMOS integration of III-V semiconductors is challenging\cite{Sze2006}. Graphene is CMOS-compatible\cite{ThomNE2018} but, in order to technologically bridge the gap with III-V-based OEMs, the bandwidth and conversion efficiency need to be improved\cite{MontNL2016}.

Here, we present a novel scheme to obtain optoelectronic mixing in graphene, employing a high-frequency graphene field effect transistor (GFET). The electrical signal is injected into the gate electrode, while a 67GHz modulated laser beam illuminates part of the graphene channel. The photocurrent contains the mixing of the electrical and optical signals. This device can be employed as optoelectronic mixer up to 67GHz, with electrical bandwidth$\sim$19.7GHz and conversion efficiency -67dB. These results improve the conversion efficiency and bandwidth of previous graphene-based OEMs\cite{MontNL2016}, paving the way for the use of graphene in CMOS-compatible OEMs.
\section{Results and discussion}
\begin{figure}
\centerline{\includegraphics[width=90mm]{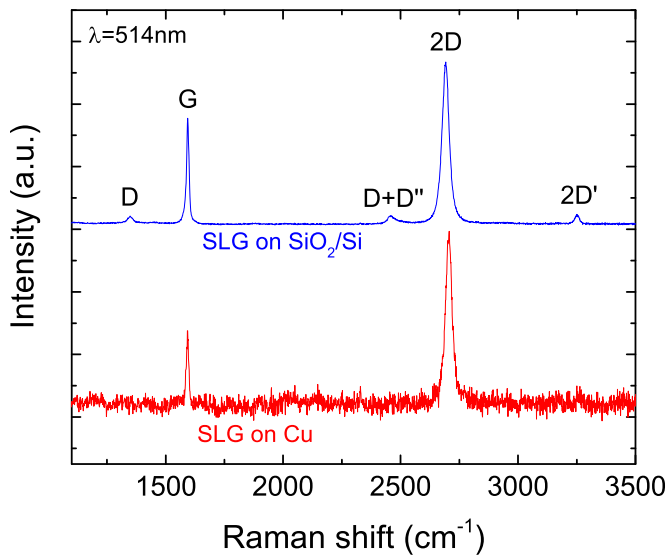}}
\caption{Representative Raman spectra at 514nm of single layer graphene (SLG) as-grown on copper (red line),and after transfer on SiO$_2$/Si substrate (blue line).}
\label{Fig_Raman}
\end{figure}
\begin{figure*}
\includegraphics[width=180mm]{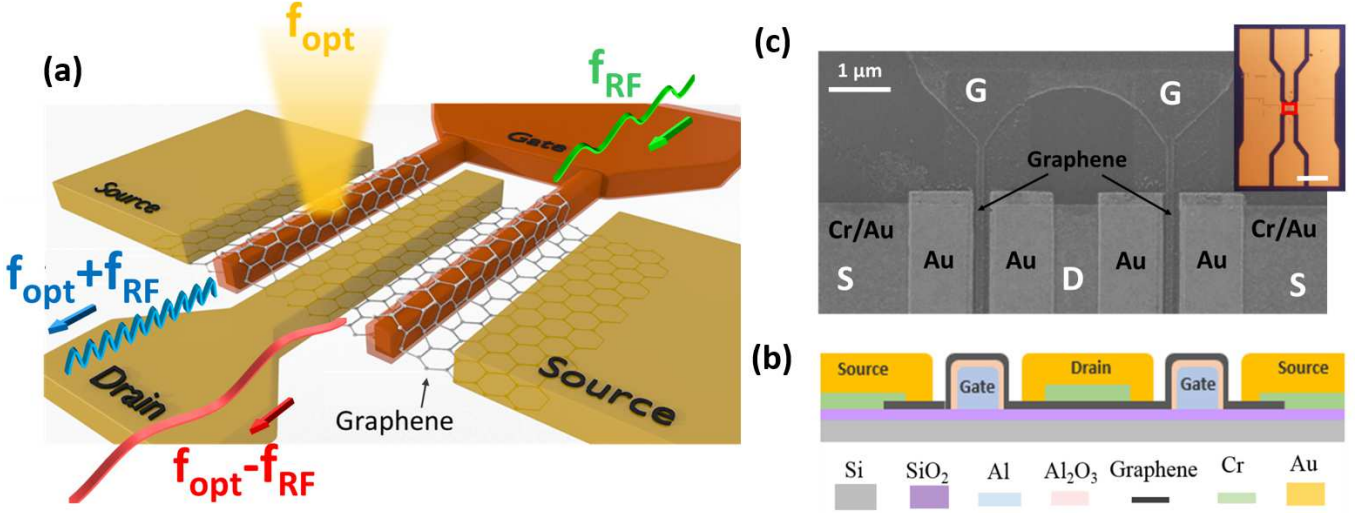}
\caption{a) Principle of operation of our optoelectronic mixer. The mixing of the electrical signal at frequency $f_{RF}$ with the photodetected signal at frequency $f_{opt}$ generates two signals at the output (drain), at different frequencies: $f_{opt}$+$f_{RF}$ and $f_{opt}$-$f_{RF}$. b) Schematic GFET cross section. The two aluminum gates (pale blue) are covered by a thin Al$_2$O$_3$ oxide (pink). SLG (black) is placed on the two gates. The drain and source contacts are realized using Au (yellow) and Cr (green). c) SEM image of GFET with double gate covered by SLG. The metal in contact with SLG is Au. The inset shows the GFET (marked by a red rectangle) integrated in a coplanar waveguide transmission line. Scale bar: 100$\mu$m}
\label{Device}
\end{figure*}
Single layer graphene (SLG) is grown via CVD on 35$\mu$m-thick Cu foil, following Ref.\cite{LiS2009}. The temperature is raised to 1000$^{\circ}$C in a H$_2$ atmosphere ($\sim$200mTorr), and kept constant for 30mins. 5sccm CH$_4$ are then added to the 20sccm H$_2$ flow to start growth, for additional 30mins at$\sim$300mTorr. The sample is then cooled at$\sim$1mTorr to room temperature. We use Raman spectroscopy at 514nm to characterize the material. Fig.\ref{Fig_Raman} shows the Raman spectrum on Cu (red line), after Cu photoluminescence removal\cite{LagaAPL2013}. The D peak is absent, indicating negligible defects\cite{FerrNN2013,CancNL2011}. The 2D peak at$\sim$2705cm$^{-1}$ is a single Lorentzian with full width at half maximum (FWHM)$\sim$31cm$^{-1}$, a fingerprint of SLG\cite{FerrPRL2006}. The G peak$\sim$1593cm$^{-1}$ has FWHM(G)$\sim$12cm$^{-1}$. The 2D to G intensity and area ratios are I(2D)/I(G)$\sim$2.4 and A(2D)/A(G)$\sim$6.3.

To prevent ohmic losses at microwave frequencies, a high resistivity Si wafer ($>$8000$\Omega$cm) covered with 300nm SiO$_2$ is used. SLG is wet transferred\cite{BaeNN2010,BonaMT2012} on it as follows. A poly(methyl methacrylate) (PMMA) layer is spin-coated on the surface of SLG/Cu and then placed in a solution of ammonium persulfate (APS) and deionized (DI) water for Cu etching\cite{BaeNN2010}. The PMMA membrane with attached SLG is then immersed into a beaker filled with DI water for cleaning APS residuals. Finally the PMMA/SLG stack is transferred onto the target substrate and the PMMA layer removed. We then characterize again via Raman spectroscopy the transferred SLG (blue curve, Fig.\ref{Fig_Raman}). The positions of the G and 2D peaks are Pos(G)$\sim$1594cm$^{-1}$ and Pos(2D)$\sim$2691cm$^{-1}$. FWHM(G) and FWHM(2D) are$\sim$11 and$\sim$34cm$^{-1}$; I(2D)/I(G)$\sim$1.6 and A(2D)/A(G)$\sim$4.5. This indicates p-doping$\sim$300meV\cite{DasNN2008,BaskPRB2009}. I(D)/I(G)$\sim$0.09, corresponding to a defect density$\sim$4$\cdot$10 $^{10}$cm$^{-2}$\cite{CancNL2011,BrunACS2014}. The SLG is then ion etched to define the channel.

Fig.\ref{Device}a is a sketch of our SLG optoelectronic mixer and illustrates its operational principle. It consists of a GFET with a double-bottom gate. A laser beam is modulated at a frequency $f_{opt}$ and focused on the GFET channel. As a result, a photocurrent that contains an AC component at frequency $f_{opt}$ flows through the SLG channel. If a radio frequency signal $f_{RF}$ is applied to the gate, the output current presents a term at frequency $f_{RF}$. When both (optical and electric) signals are applied, the device acts as an optoelectronic mixer: the output contains the product of the two signals, and two AC components at $f_{opt}$+$f_{RF}$ and $f_{opt}$-$f_{RF}$ appear.

A schematic cross section of the bottom gate GFET is in Fig.\ref{Device}b. The fabrication starts by patterning the double bottom-gates by e-beam lithography (EBPG 5000 Plus). The gates are made of a 40nm thick Al layer deposited by evaporation. A 4nm Al$_{2}$O$_{3}$ layer is formed on top of the gates by exposing the substrate to pure oxygen for 30mins\cite{WeiIEEE2015} with a Oxford Plasmalab80Plus at an oxygen pressure$\sim$100mTorr. This thin oxide acts as gate dielectric. The source and drain contacts are made in a two-steps process. First, Cr/Au (5/50 nm) pre-contacts are deposited on SLG. Then, ohmic contacts are obtained by placing 30nm Au on the Cr/Au-SLG junction. Finally, a coplanar waveguide is built with a Ni/Au film (50/300nm). Fig.\ref{Device}c is a scanning electron microscopy (SEM) image of the bottom gates covered by SLG. The inset shows a GFET integrated in the coplanar waveguide. The red square indicates the area occupied by the GFET. The bottom gate transistor design is suitable for optoelectronic mixing since 1) the SLG channel is on the gate and can be directly illuminated; 2) the use of a thin (4nm) Al$_2$O$_3$ dielectric and short gate($<$0.4$\mu$m or less) ensures high frequency operation\cite{SchwN2015,GuoNL2013,Poza2009}.

The device exhibits a cut-off frequency $f_t\sim$25GHz and a maximum oscillating frequency $f_{max}\sim$14GHz, as deduced from the S-parameters measured with a VNA Network Analyzer (Agilent, E8361A). To calibrate the VNA, we use the Line-Reflect-Reflect-Match (LRRM) calibration\cite{DaviIEEE1990}. This allows us to eliminate the error in the S-measurements introduced by the environment, such as the cables and  the probe tips used to electrically contact the device under test, and the VNA non-idealities.

\begin{figure}
\centerline{\includegraphics[width=85mm]{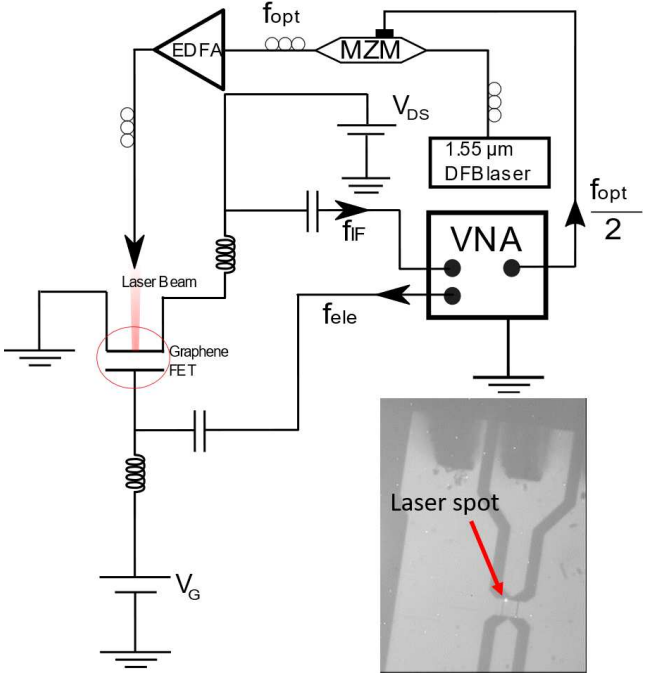}}
\caption{Measurement setup. A CW laser is modulated via a MZM. It is then amplified with an EDFA and focused on the GFET. An AC signal is applied to the gate. The output (IF) is measured on a VNA. Inset: optical image of device with laser focused on the channel.}
\label{experiment}
\end{figure}

The set-up in Fig.\ref{experiment} is used to measure photocurrent and optoelectronic mixing. The output of a 1.55$\mu$m distributed feedback (DFB) laser is modulated by a Mach Zehnder modulator (MZM) in the double sideband suppression carrier (DSB-SC) mode\cite{InagIEICE2012}, to obtain a modulated beam at $f_{opt}$. This is then amplified with an Erbium-doped fiber amplifier (EDFA). The maximum $f_{opt}$ that our setup can probe is 67GHz. The diameter of the focused laser spot is$\sim$2$\mu$m (inset of Fig.\ref{experiment}). The maximum power impinging on the sample is 60mW, which corresponds to $\sim$20mW/$\mu$m$^2$. The gate and drain are connected to a vector network analyzer (VNA) with two high-frequency (67GHz) air coplanar probes. Bias tees are used to add a DC bias to channel and gate electrode.

\begin{figure}
\centerline{\includegraphics[width=90mm]{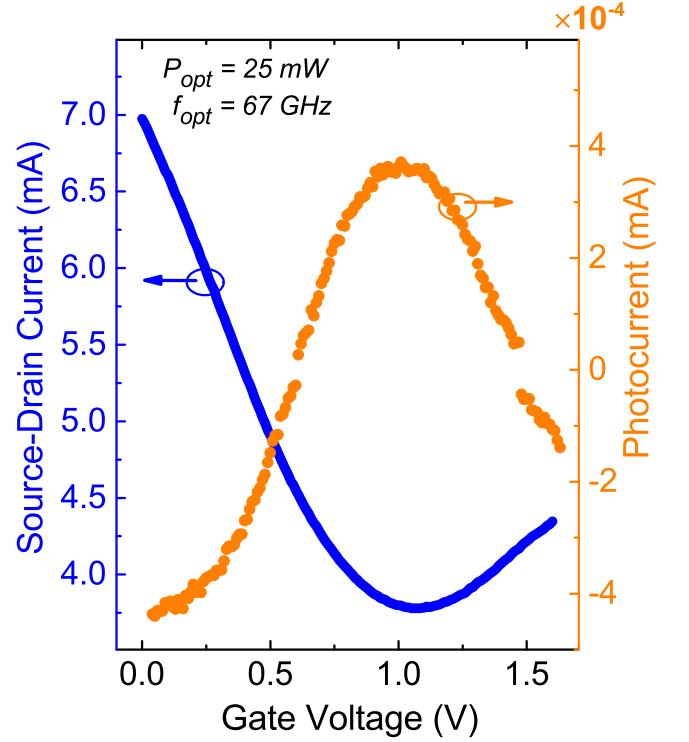}}
\caption{Blue curve: Source drain current for $V_{DS}$=200mV. Orange curve: photocurrent generated by a 25mW beam focused on the SLG channel.}
\label{photocurrent}
\end{figure}

We now present the results on a representative device with SLG channel width and length W=24$\mu m$ and L=400nm and gate length $L_{G}$=200nm. The blue curve in Fig.\ref{photocurrent} is the source-drain current as a function of gate voltage, at $V_{DS}$=200mV. The minimum conductance is reached at a gate-source bias $V_{GS}$=1.1V, which corresponds to the charge neutrality point voltage ($V_{CNP}$). The field effect mobility $\mu$ is calculated as $\mu$=L$_G$g$_m$/W$\cdot$C$_G$V$_{DS}$\cite{LemmIEEE2007}. The transconductance g$_m$= \textit{d}$I_{DS}$/\textit{d}$V_{GS}$\cite{Sze2006} is obtained from the transfer characteristic $I_{DS}$($V_{GS}$) at V$_{DS}$=10mV. The gate capacitance C$_G$ is extracted from the S-parameters measurement. We get $\mu_{fe}\sim$3800cm$^2$V$^{-1}$s$^{-1}$, consistent with room-temperature $\mu$ in non-encapsulated CVD SLG\cite{LiS2009}.

We first consider the photoresponse. The device is biased at $V_{DS}$=200mV and illuminated with a laser modulated at $f_{opt}$=67GHz. The electrical power $P_{RF}$ measured by the VNA is used to derive the photocurrent $I_{ph}$. From Joule's law\cite{Sze2006} $I_{ph}=\sqrt{\frac{P_{RF}}{Z_{VNA}}}$, with $Z_{VNA}=50\Omega$ the VNA input impedance. The photocurrent as a function of $V_{GS}$ is the red curve in Fig.\ref{photocurrent}, for 25mW incident power. The photocurrent sign depends on the gate voltage. $I_{ph}$ is positive and has a local maximum close to the CNP. For carrier concentration $n>3.5\cdot10^{12}cm^{-2}$ (or $V_{GS}-V_{CNP}>0.5V$), $I_{ph}$ becomes negative. This is typical of biased SLG PDs\cite{KoppNN2014,FreiNP2013}. For $n<3.5\cdot10^{12}cm^{-2}$, the laser power increases the charge carrier density (photoconductive regime) and the channel conductance increases. Therefore, the photocurrent has the same sign as the DC current in the channel due to the DC bias. At high doping ($n>3.5\cdot10^{12}cm^{-2}$) the sign of the photocurrent is opposite to the DC current, due to a decrease of $\mu_{fe}$ resulting from an increase of the carrier temperature caused by the laser (bolometric regime)\cite{FreiNP2013,FerrN2015,SassNC2017}.

\begin{figure*}
\centerline{\includegraphics[width=175mm]{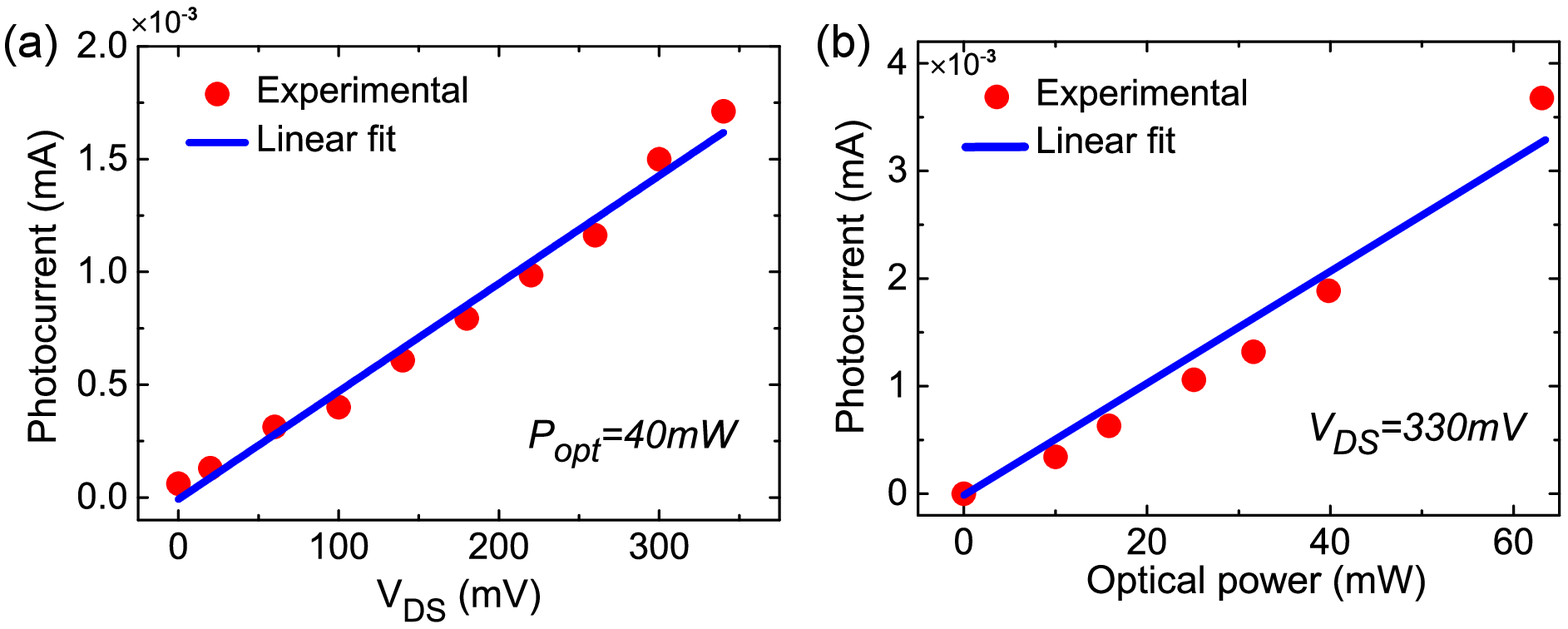}}
\caption{Blue curve: Source drain current for $V_{DS}$=200mV. Red curve: photocurrent generated by a 25mW beam focused on the SLG channel. a) Photocurrent as a function of $V_{DS}$ at 40mW. b) Photocurrent as a function of optical power at $V_{DS}$=330mV}
\label{VD}
\end{figure*}

\begin{figure}
\centerline{\includegraphics[width=80mm]{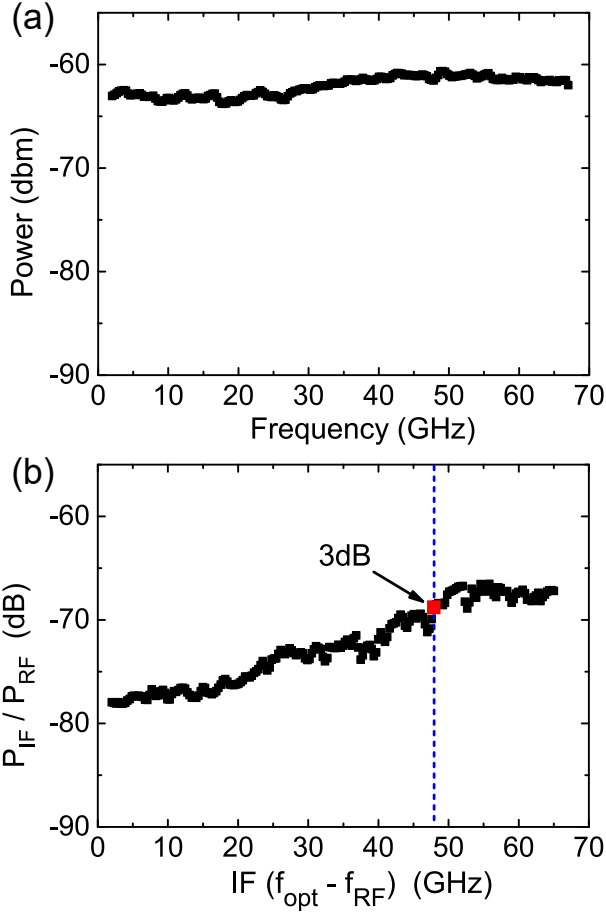}}
\caption{(a) Maximum photodetected power at $V_{GS}$=0V, $V_{DS}$=330mV, as a function of $f_{opt}$. (b) Downconversion efficiency at $V_{GS}$=0.6V, $V_{DS}$=330mV. In both plots, the optical power is 60mW.}	
\label{MixMax}
\end{figure}

The external photoresponsivity $R_{ext}$ is defined as\cite{KoppNN2014,DeFaACS2016} R$_{ext}$=$\frac{|I_{ph}|}{P_{cw_{eff}}}$. Here, $P_{cw_{eff}}=31\%P_{cw}$ is the fraction of the optical power coupled to the SLG channel. We get $R_{ext}\sim$0.22mA/W. For $V_{GS}$=0 V, the device reaches its maximum $I_{ph}\sim$-4.2$\cdot$10$^{-4}$mA. At this $V_{GS}$, the photocurrent generated by a 67GHz laser modulation is measured as a function of DC bias and optical power. Figs.\ref{VD}a,b plot the photocurrent as a function of DC bias at P$_{opt}$=40mW and as a function of the the optical power for $V_{DS}$=330mV. The response is linear in both cases, as expected for a photoconductor\cite{FreiNP2013}. The frequency response of the photodetected power is then measured as a function of $f_{opt}$, Fig.\ref{MixMax}a. We get a flat response over the whole band that can be investigated by our VNA, showing that the intrinsic photodetection bandwidth is$>$67GHz.

In order to operate the device as an optoelectronic mixer (instead of a PD), an RF signal at a frequency $f_{RF}$ is added to the DC gate, (Fig.\ref{Device}a). $f_{opt}$ is maintained at 67GHz, while $f_{RF}$ is swept between 2 and 65GHz. A VNA is used to record $P_{IF}$, the transistor power at the intermediate frequency $f_{IF}=f_{opt}-f_{RF}$.

An important parameter for optoelectronic mixers is the downconversion efficiency\cite{Poza2009}: $P_{IF}/P_{RF}$[dB], with $P_{RF}$ the power at the source and $P_{IF}$ that measured at the VNA. For this device, the maximum downconversion efficiency is -67dB at $V_{GS}$=0.6V. For this $V_{GS}$, Fig.\ref{MixMax}b plots the downconversion efficiency as a function of $f_{IF}$. The device has a 3dB bandwidth$\sim$19.7GHz. The downconversion efficiency is 21dB higher than Ref.\cite{MontNL2016}, where SLG mixers operating at 0-30GHz were reported.

\begin{figure*}
\centerline{\includegraphics[width=175mm]{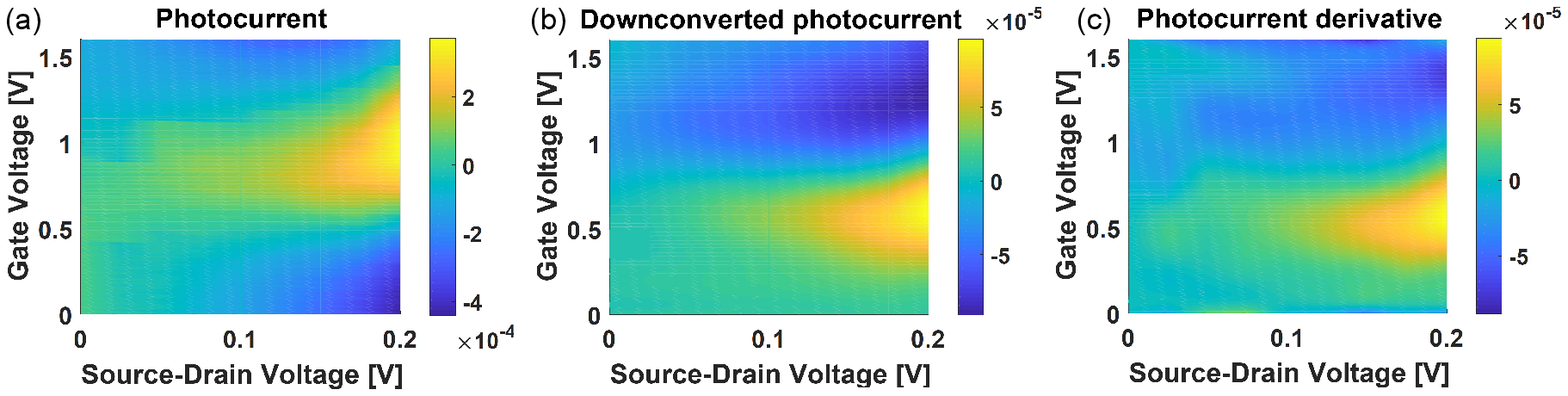}}
\caption{a) Photocurrent map as a function of $V_{GS}$, $V_{DS}$. b) Downconverted photocurrent map as a function of $V_{GS}$, $V_{DS}$. c) Derivative of a) with respect to $V_{GS}$. The photocurrent values are expressed in mA.}
\label{map}
\end{figure*}

\begin{figure}
\centerline{\includegraphics[width=85mm]{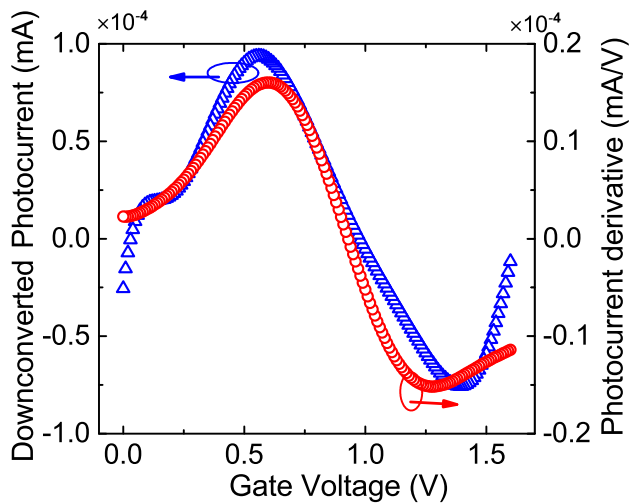}}
\caption{Red curve: cut of Fig.\ref{map}c for $V_{DS}$=200mV. Blue curve: cut of Fig.\ref{map}b for $V_{DS}$=200mV}.
\label{slice}
\end{figure}

Fig.\ref{map}a is a color map of the 67GHz photocurrent as a function of $V_{GS}$, $V_{DS}$. We then add to the DC gate bias an electrical signal at 10GHz. The resulting downconverted photocurrent at $f_{IF}$=57GHz is plotted as a function of $V_{DS}$, $V_{GS}$ in Fig.\ref{map}b. By differentiating the map in Fig.\ref{map}a with respect to $V_{GS}$, we obtain Fig.\ref{map}c, which resembles Fig.\ref{map}b. This is best seen in Fig.\ref{slice}, which plots both values as a function of $V_{GS}$ for $V_{DS}$=200mV. The curves of the downconverted photocurrent and of the derivative of the photocurrent can be superposed.

This can be explained by carrying out a small-signal analysis. Let us consider the modulated optical power impinging on the PD, $P_{opt}=P_{cw}+P_{mod}sin(2\pi f_{opt} t)$, with $P_{mod}$ the amplitude of the varying part of the optical power. The generated photocurrent is proportional to $P_{opt}$ through the factor $R_{ext}$. This depends on $V_{GS}$, as for Fig.\ref{photocurrent}, and is almost independent on $f_{opt}$, Fig.\ref{MixMax}b. Therefore, the photocurrent can be written as:
\begin{equation}\label{eq:photo}
I_{ph}(V_{GS})= R_{ext}(V_{GS})[P_{cw}+ P_{mod}sin(2 \pi f_{opt} t)]
\end{equation}
By applying to the gate a DC bias $\bar{V}_{GS}$ and a small signal $\delta_{V_{GS}}sin(2 \pi f_{RF} t)$, we can write:
\begin{equation}\label{eq:photo2}
R_{ext}(V_{GS})= R_{ext_{DC}}(\bar{V}_{GS}) + \delta_{V_{GS}}\Delta_{R_{ext}}sin(2 \pi f_{RF} t)
\end{equation}
where
\begin{equation}\label{eq:photo3}
\Delta_{R_{ext}}= \beta({f_{RF}})\frac{d R_{ext}(V_{GS})}{dV_{GS}}\Bigr|_{V_{GS}=\bar{V}_{GS}}
\end{equation}
We include a dependence on injected electrical frequency through a frequency-dependent proportionality constant $\beta({f_{RF}})$. The total photocurrent has 4 terms:
\begin{equation}\label{eq:photo4}
\begin{aligned}
& I_{ph}=R_{ext_{DC}}(\bar{V}_{GS})P_{cw}+\\
& \delta_{V_{GS}}\Delta_{R_{ext}}P_{cw}sin(2 \pi f_{RF} t)+\\
& R_{ext_{DC}}(\bar{V}_{GS})P_{mod}sin(2 \pi f_{opt} t)+\\
& \delta_{V_{GS}}\Delta_{R_{ext}}P_{mod}sin(2 \pi f_{RF} t)sin(2 \pi f_{opt} t)
\end{aligned}
\end{equation}
The first is the DC photocurrent. The second describes the DC photocurrent modulated by the electrical signal. The third represents the photocurrent modulated at $f_{opt}$, Fig.\ref{photocurrent}. The fourth describes the optoelectronic mixing and can be rewritten as:
\begin{equation}\label{eq:mixpart}
\begin{aligned}
& \delta_{V_{GS}}\Delta_{R_{ext}}P_{mod}sin(2 \pi f_{RF} t)sin(2 \pi f_{opt} t)=\\
& {\frac{1}{2}} \delta_{V_{GS}}\Delta_{R_{ext}}P_{mod}\{cos[2 \pi (f_{RF}-f_{opt})t]+
\\
& -cos[2\pi (f_{RF}+f_{opt})t)\}
\end{aligned}
\end{equation}

Eq.\ref{eq:mixpart} has two components at frequencies $f_{opt}$+$f_{RF}$ and at the intermediate frequency $f_{IF}=f_{opt}$-$f_{RF}$. It shows that the mixed signal depends exclusively on $\Delta_{R_{ext}}$, i.e. on derivative of $R_{ext}$ with respect to $V_{GS}$, not on $R_{ext}$ itself, in accordance with Fig.\ref{map}. $\Delta_{R_{ext}}$ is maximum for $V_{GS}\sim$0.6V,Fig.\ref{slice}, i.e. in a region where the photocurrent changes sign, Fig.\ref{photocurrent}. The sharper is the transition between the two competing phenomena (photoconductive and bolometric) generating the photocurrent, the higher is $\Delta_{R_{ext}}$ and the optoelectronic mixing efficiency. The control of the transition between the two different photocurrents could lead to a maximization of $\Delta_{R_{ext}}$ and, in turn, to a maximization of device performances.

The 3dB bandwidth is$\sim$19.7GHz when operated as a optoelectronic mixer, Fig.\ref{MixMax}. This behavior is modeled in Eq.\ref{eq:photo3} by including the factor $\beta({f_{RF}})$. To understand the optoelectronic mixing dependence on $f_{RF}$ and, thus, on $f_{IF}$ ($f_{opt}$ being fixed), one may consider a typical figure of merit of high frequency transistors: the transducer power gain\cite{Poza2009}, defined as: G$_T$=$\frac{P_{load}}{P_{avs}}$. $P_{load}$ is the power delivered to the load and $P_{avs}$ is the source power. $G_T$ coincides with the modulus of the S$_{21}$ parameter when source and load are matched. This is the case in our measurements, where the power is delivered from the VNA 50$\Omega$-source and measured on a 50$\Omega$ receiver. The measured downconversion efficiency (i.e. the transducer power gain $G_T$\cite{Poza2009}) is close to the S$_{21}$ parameters. An external impedance matching could in turn enhance the efficiency and increase downconversion efficiency by maximizing the power delivered by the GFET.

We do not observe saturation in the photodetected signal at the highest power available in our setup. Thus, illuminating a wider channel surface while maintaining the same optical power density should increase the downconversion efficiency. Impedance matching can also be used to increase both bandwidth and efficiency.
\section{Conclusions}
We reported a GFET operating as an optoelectronic mixer for frequencies up to at least 67GHz. The photodetection bandwidth exceeds 67GHz. The bandwidth of the device operated as an optoelectronic mixer is 19.7GHz. A model was presented to describe the measured downconversion efficiency. These results pave the way for the use of graphene-based transistors as optoelectronic mixers in applications exploiting mm-waves, such as telecommunications and RADAR/LIDARs.
\section{Acknowledgements}
We acknowledge funding from EU Graphene Flagship, the French RENATECH network, ERC Grants Hetero2D and MINERGRACE, EPSRC Grants EP/K01711X/1, EP/K017144/1, EP/N010345/1, EP/L016087/1.

\end{document}